# Tailoring chaotic motion of microcavity photons in ray and wave dynamics by tuning the curvature of space

Wei Lin [1][††], Yechun Ding[1][††], Yongsheng Wang[1,2], Yanpeng Zhang[1], Feng Yun[1,3] and Feng Li[1,3*]

[1] *Key Laboratory for Physical Electronics and Devices of the Ministry of Education & Shaanxi Key Lab of Information Photonic Technique, School of Electronic Science and Engineering, Faculty of Electronic and Information Engineering, Xi'an Jiaotong University, Xi'an 710049, China;*
[2] *Department of Electrical and Photonics Engineering, Technical University of Denmark, Kgs. Lyngby, 2800, Denmark;*
[3] *Solid-State Lighting Engineering Research Center, Xi'an Jiaotong University, Xi'an 710049, China*

Microcavity photon dynamics in curved space is an emerging interesting area at the crossing point of nanophotonics, chaotic science and non-Euclidean geometry. We report the sharp difference between the regular and chaotic motions of cavity photons subjected to the varying space curvature. While the island modes of regular motion rise in the phase diagram in the curved space, the chaotic modes show special mechanisms to adapt to the space curvature, including the fast diffusion of ray dynamics, and the localization and hybridization of the Husimi wavepackets among different periodic orbits. These observations are unique effects enabled by the combination of the chaotic trajectory, the wave nature of light and the non-Euclidean orbital motion, and therefore make the system a versatile optical simulator for chaotic science under quantum mechanics in curved space-time.

**microcavity, chaotic motion, curved space**

## 1 Introduction

Optical microcavities supporting Whispering-Gallery mode (WGM) have been widely investigated for decades, leading to emerging applications such as on-chip micro-lasers [1-3], frequency combs [4,5], isolators [6,7], environmental sensors [8,9], and enhanced light-matter interaction [10]. The optical confinement of the WGMs is provided by the total internal reflection (TIR) at the cavity boundaries, resulting in rich photon dynamics engineered by the delicately designed shapes of the cavity boundary. In particular, when the boundary of the cavity is slightly deformed to break the rotational symmetry, chaotic motion of photons can arise leading to exotic dynamic properties [11-13], with the Limaçon [14,15] and the Face [16,17] cavities as well-known examples. The photon trajectories in such cavities, when analyzed with ray optics, exhibit classical chaotic nature usually characterized by the coexistence of the WGMs, the periodic island modes and the chaotic modes which are visualized in the phase diagram called Poincaré surface of section (PSOS) [17-19]. Nevertheless, when the wave nature of the light is taken into consideration, the uncertainty relation between the position and the momentum of the cavity photon plays a non-trivial role, transforming the system to an optical simulator of chaotic phenomena incorporating quantum properties [13,20]. Herein the terms "classical" and "quantum" are optical analogs to mass-point-like particles and matter waves, respectively, and should be distinguished from the concepts of classical and quantum lights. In such a scenario, the chaotic modes can be weakly coupled to the WGMs, forming channels for dynamic tunneling [21]. This mechanism has been widely applied to achieve unidirectional lasing emission [22-24], broadband momentum transformation [12,25], broadband frequency combs [26,27] and photon transport regulation [28].

While the shape engineering of WGM cavities tailors the photon dynamics in a two-dimensional (2D) plane, three-dimensional (3D) control of photon dynamics is simultaneously investigated using spatially curved structures, with rich physics and potential applications revealed in curved photonic circuits [29], self-rolled-up microtubes [30,31] and microdisks [32,33], geodesic lenses [34,35] and Möbius strip microlasers [36]. Such curved 3D photonic devices would potentially increase the capacity of the integrated photonic chips by working with higher dimension and more degrees of freedom. Very recently, the chaotic behaviors of photons in a WGM microcavity were studied in curved space [37]. It is shown that the space curvature reduces the dissipative loss of the low-periodic island modes which basically exhibit regular motion. Nevertheless, it is not clear yet how the chaotic motion reacts to the varying space curvature, and therefore, a comprehensive understanding of the different photon dynamics between the regular and chaotic motions remains unexplored. Especially, it is interesting to know whether there exist unique quantum effects (analog of wave optics) associated with the space curvature, which can neither be predicted via the classical picture of ray dynamics nor be supported by flat cavities.

---

*Corresponding author (email: felix831204@xjtu.edu.cn)
††These authors contributed equally to this work.

In this Article, we report on the sharp difference between the regular and chaotic photon dynamics in an asymmetric WGM microcavity subjected to the varying space curvature, based on both ray dynamics and wave optics. We show that unlike the regular eigenmodes which rise in phase space with increasing space curvature, the chaotic eigenmodes show special mechanisms to adapt to the space curvature, including localization and hybridization of the Husimi wavepackets among different periodic orbits. We demonstrate that these observations are unique effects combing the wave nature of light, the chaotic property of motion, and the curved trajectories with non-Euclidean geometry, which contribute significantly to the widely-investigated area of chaotic motions under the laws of quantum mechanics. Meanwhile, our studies also provide novel principles for the design of optoelectronic devices based on the chaotic motion of microcavity photons which can be controlled at higher dimensions.

## 2 The model of the curved microcavity

The curved WGM microcavity with a near-circular boundary is defined on a spherically-curved surface, as illustrated in Figure 1 (a), in which ROC denotes the radius of curvature of the spherical surface, and ROB denotes the radius of the boundary of the cavity measured by the geodesic distance from the cavity center O' [37]. Photons travel along the geodesic lines of the spherical surface and get reflected at the cavity boundary. The effective space curvature $k$=ROB/ROC defines the geometric shape of the spherical surface regardless of the cavity size and therefore determines the ray dynamics [37]. An example of the calculated light trajectories

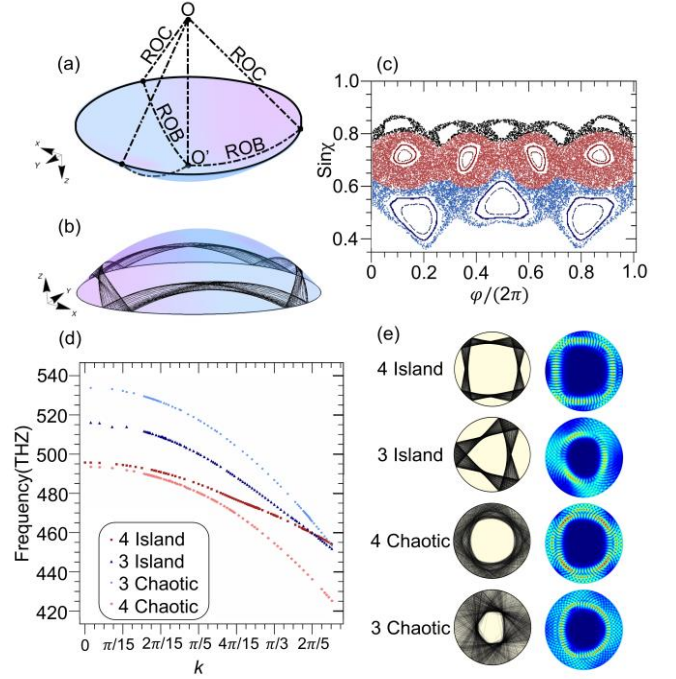

**Figure 1 Basic optical properties of the asymmetric microcavity on curved surface.** (a) Schematic picture of the microcavity. O and O' are the center of the spherical surface and the geometric center of the cavity, respectively. ROC: radius of curvature; ROB: radius of the boundary. (b) Light trajectories of a symmetric circular cavity with ROB=0.8 defined on a curved surface of ROC=1. (c) PSOS of a flat Face microcavity. Four distinct regions are depicted with different colors: deep red and light red for four-period island and chaotic modes; deep blue and light blue for three-period island and chaotic modes; black for five-period modes. (d) The eigen-frequencies of the three and four period island and chaotic modes as a function of $k$. (e) Light trajectories (left panels) and optical field spatial profiles (right panels) of each mode in (d) (top-down view).

in a symmetric circular cavity with $k$= 0.8 (ROC = 1, ROB = 0.8) is shown in Fig.1 (b). To incorporate chaotic motion, an asymmetric boundary is defined via the well-known "Face cavity", same as in Ref. [37]:

$$R = \begin{cases} R_0 \cdot (1+a_2 \cdot \cos^2 \phi + a_3 \cdot \cos^3 \phi) & \phi \in \left(\frac{-\pi}{2}, \frac{\pi}{2}\right) \\ R_0 \cdot (1+b_2 \cdot \cos^2 \phi + b_3 \cdot \cos^3 \phi) & \phi \in \left(\frac{\pi}{2}, \frac{3\pi}{2}\right) \end{cases} \quad (1)$$

in which $R$ denotes the ROB measured by the geodesic distance. $\phi$ is the azimuthal angle, and $R_0$, $a_2$, $a_3$, $b_2$, and $b_3$ are the boundary shape parameters, which are chosen to be $R_0$ = 1, $a_2$ = -0.1329, $a_3$ = 0.0948, $b_2$ = -0.0642, $b_3$ = -0.0224 in our studies. Fig. 1 (c) presents the calculated PSOS for a flat microcavity ($k$=0). We focus on the island motions of the 5-period, 4-period and 3-period orbits (referred to as island mode hereafter) as well as the chaotic motions mostly around these islands (referred to as chaotic mode hereafter), which are shown in black, red and blue colors. To be concise, other periodic orbits of trajectories are not drawn in the PSOS. Correspondingly, we perform simulations with wave optics using

COMSOL Multiphysics (Version 5.6). The cavity surface is set to be a thin layer with a thickness d =50 nm and refractive index $n = \sqrt{10}$ surrounded by air [37]. When increasing the effective curvature $k$, the light trajectories, as well as the optical fields, are pushed to the cavity periphery and therefore the resonant frequencies of the modes change, as shown in Fig. 1(d). The four presented modes, whose field distributions fall mainly in the 3-period and 4-period island and chaotic regions and exhibit no crossing points in most of the studied range of $k$, are intentionally chosen to allow independent analysis of the evolution of each mode. The light trajectories and optical field distribution of the four modes at $k=0$ are shown in the left and right panels of Fig. 1(e), in which the obvious correspondences of spatial features are seen between the ray trajectories and the optical fields.

## 3 Regular and chaotic photon motions analyzed with ray dynamics

The general mechanism of the interaction between the space curvature and photon dynamics has been previously investigated [37]. Briefly speaking, for the orbits of the same order of period, the reflection angle increases with increasing $k$, owing to the simple fact that inner angles of polygons generally get larger on a spherically curved surface. As a result, the phase diagram in the PSOS experiences an upward shift with increasing $k$, from which a global rise of the quality factors (Q-factors) of the optical modes is expected [37]. We start the investigation by calculating the Q-factors (done with wave optics simulations) as a function of $k$ for the four eigenmodes in Fig. 1(d), as plotted in Figure 2(a). It is obvious that the island modes exhibit a remarkable rise of Q-factor with increasing $k$ as expected, while the dips at certain discrete $k$ values correspond to the coupling with coexisting low-Q states and are thereby not an intrinsic property of the island modes [37]. Meanwhile, surprisingly, the Q-factors of the chaotic modes do not increase with $k$ but, on the contrary, even decreases. As well explained previously, reaching a sufficiently high position in the PSOS is crucial to form a high-Q mode [37]. Therefore, this observation raises the question whether the chaotic modes around the islands really rise together with the island, or the rising chaotic sea with space curvature (a good example can be found in Figure S2 of the supplementary information of Ref [37]) is in fact an illusion.

To make this point clear, we investigate the formation process of the chaotic motion by ray dynamics, with the results presented in Figs. 2(b) to 2(g), which include three situations corresponding to small ($k=\pi/100$), intermediate ($k=\pi/5$) and large ($k=\pi/3$) space curvatures. For each space curvature, we start a ray tracing from a point belonging to the 4-period island mode for understanding the regular motion, and another ray tracing from another point at the same range of height of the 4-period island mode but inside the chaotic sea for understanding the chaotic motion, with light trajectories up to 10000 times of reflections recorded in both tracings. The results show that the island mode always stays on the islands during the entire process, exhibiting high stability for all space curvatures, regardless of the size and the shape of the islands, as demonstrated in Figs. 2(b), 2(d) and 2(f). On the contrary, the behavior of the chaotic motion differs dramatically among different space curvatures. Although it is expected that the motion will finally diffuse in the phase diagram after a sufficiently long process of propagations and reflections, the speed of the diffusion matters. Statistically, if the motion stays a sufficiently long time surrounding the 4-period region before diffusing out, a well-defined chaotic mode with a quasi-4-period orbit can form, even though suffering higher loss than the island modes. On the other hand, if the motion diffuses too fast, an optical mode can hardly form. It is clearly seen from the results that the diffusing speed presents a crucial increase with space curvature. While at low space curvature the motion keeps to stay in the 4-period region up to ~5000 reflections before diffusing to the 3- and 5-period regions (Fig. 2(c)), the persistence drops to ~3500 reflections for intermediate space curvature (Fig. 2(e)) and to only ~1100 reflections for large space curvature (Fig. 2(g)). This unambiguously indicates that the chaotic motions of cavity photons react to the variation of space curvature in an essentially different way compared to the regular motions.

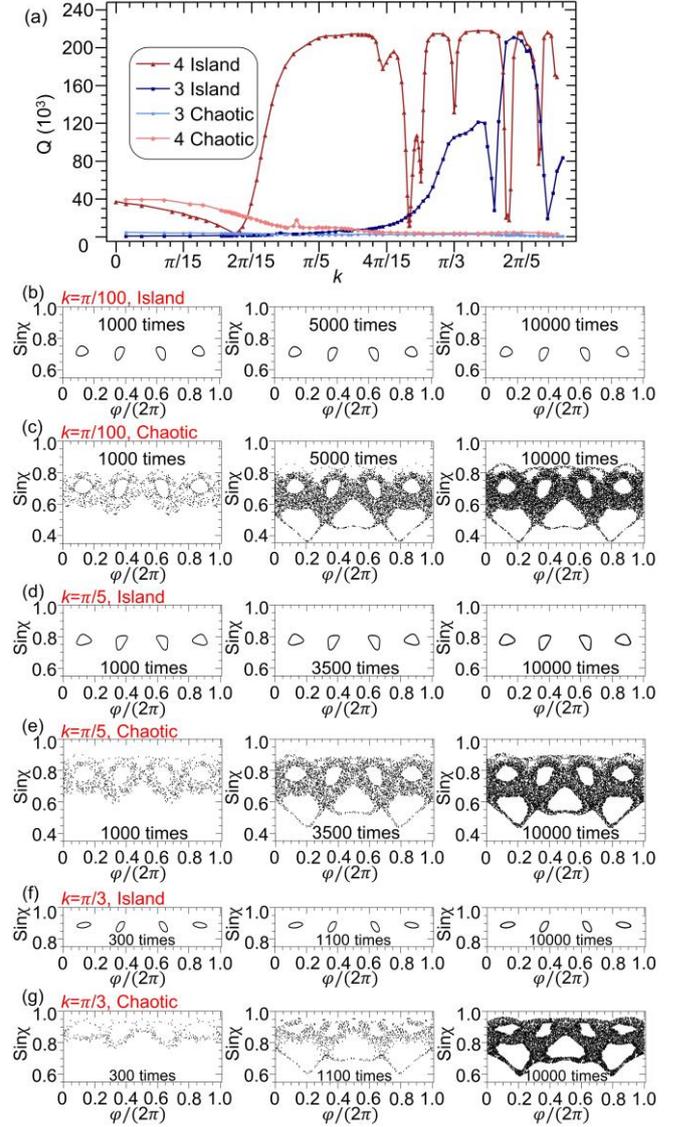

**Figure 2** The analysis of the regular and chaotic motions with ray dynamics. (a)The Q-factors as a function of $k$ for the 4-period island mode (deep red), 4-period chaotic mode (light red), 3-period island mode (deep blue), and 3-period chaotic mode (light blue), obtained with wave optics simulations using COMSOL. (b)-(g) PSOS associated with the light trajectories for record island and chaotic motions under different curvatures: (b)(c) for $k=\pi/100$, (d)(e) for $k=\pi/5$, and (f)(g) for $k=\pi/3$. Each calculation includes 10,000 tines of reflections, and each graph is presented for a specific time of reflections.

Now we get to the point to answer the question raised before the analysis. Although the chaotic sea seems to experience a global rise together with the islands during the increase of the space curvature, its dynamics becomes much more diffusive in the PSOS so that the chaotic motion is completely reconstructed, losing its correspondence to the situations of zero and small space curvatures. Interestingly, while the space curvature serves as an effective compressive force squeezing the phase diagram towards the top of the PSOS, the motions of different natures react in different ways. The island mode, protected by the high stability of its regular motion, exhibits good adaption to the compression and exists robustly while getting squeezed in size. On the contrary, the chaotic mode does not stand the compression due to its low stability, and thereby shows diffusive features, refusing to be squeezed. It should be noted that different starting points of ray tracing may result in different diffusion speeds for a single round of calculation, nevertheless, the relation between the diffusion speed and space curvature holds statistically for many rounds of calculations with different starting points. This mechanism leads to an actual rise for the island modes but not for the chaotic modes, resulting in the large discrepancy of Q vs. $k$ shown in Fig. 2(a).

## 4 Regular and chaotic photonic modes analyzed with wave dynamics

In order to verify the conclusions drawn by ray dynamics, we perform wave optics simulation which is an analog of quantum mechanics. Herein we focus on the 4-period island and chaotic modes whose frequencies are labeled in deep red and the light red in Fig. 1(d), respectively. In order to reach a decent comparison between the ray and wave dynamics, we calculated the Husimi projections of the two modes at three different space curvatures, as shown in Figs. 3(a) to 3(f), with the ray trajectories and field distributions presented as insets. The Husimi projection was frequently used to map the field distribution obtained by wave simulation onto the PSOS for flat WGM cavities [38,39], which we find also applicable to curved spaces. It is clearly seen that when the space curvature increases, the wavepacket of the island mode in phase space rises together with the island features of ray dynamics, while the wave-packet of the chaotic mode does not rise at all. Interestingly, the shape of the wavepacket changes in a special way adapting to the rise of the PSOS. At small $k$ [Fig. 3(d)], the spatial distribution of the Husimi wavepacket exhibits four lobes which are almost discrete in the chaotic sea, presenting generally the 4-periodic feature. We label the lobes in the graph as lobes 1, 2, 3 and 4 for further discussion.

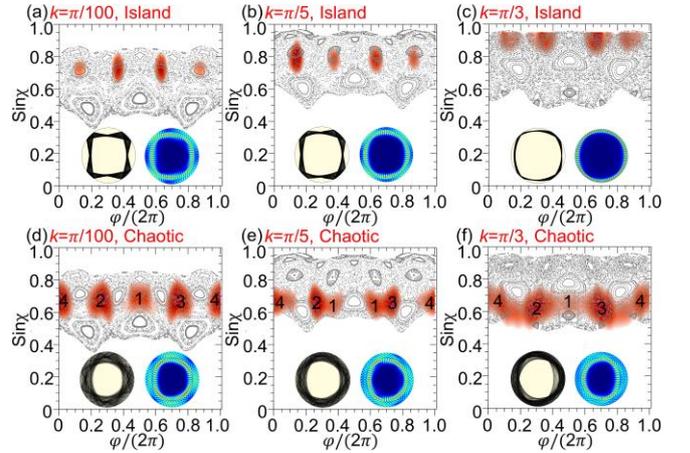

**Figure 3 The analysis of the regular and chaotic motions with wave dynamics.** (a)-(c) Husimi projections (red shade) of the 4-period island mode under a series of effective space curvatures: (a) $k=\pi/100$, (b) $k=\pi/5$, and (c) $k=\pi/3$. (d)-(f) Husimi projections of the 4-period chaotic mode under the same series of effective space curvatures. The ray trajectories and spatial field distributions of each mode are included as insets in each graph.

When $k$ increases [Fig. 3(e)], the 3-period islands rise to the height of the wavepacket of the four lobes, break the central lobe 1 into two parts, and push them apart into the chaotic sea. Meanwhile, the islands also push the lobes 2 and 3 towards the center into the chaotic sea. As a result, the two parts of the broken lobe 1 joins lobes 2 and 3 respectively, forming two lobes with three maxima. On the contrary, lobe 4 fits in a larger region of the chaotic sea and expands laterally. As a whole, the wavepacket forms a distribution of three lobes with four maxima, presenting a hybrid feature combining the 3-period and 4-period motion. When *k* further increases [Fig. 3(f)], the wavepacket meets a region absent of big islands, and thereby expands its wavepacket laterally into the chaotic sea while simultaneously trying to recover the central lobe 1, forming another type of hybrid distribution. Such a special interacting mechanism between the classical phase diagram and the quantum wavepacket is a unique effect of the curved space, which cannot be observed in flat chaotic cavities.

Additionally, we note that the wave optics yields an effect of localization in contrast with the ray dynamics. Although ray dynamics predicts that the chaotic motions diffuse faster in phase space at larger space curvatures, it does not predict the exact location of the optical eigenmodes in the PSOS, since the diffusion is directionless. The wave optics, nevertheless, indicates clearly where the wave-packet locates and how much they really spread, presented by the Husimi projections. Such a feature of localization, originated from the wave nature of light, is indispensable to understand the photon dynamics in curved space. Indeed, throughout the range of light frequencies in which the size of our simulation grid gives reasonable results, the Husimi projections of the chaotic modes never reach any higher position in the PSOS. Although it can be inferred that at much higher frequencies (therefore much shorter wavelengths) where the light waves become much more ray-like, the Husimi projection would become similar to the PSOS. However, this aspect of study is out of the scope of our work as it sheds little light on the wave nature of motion and requires a calculating power exceeding the capability of our work station.

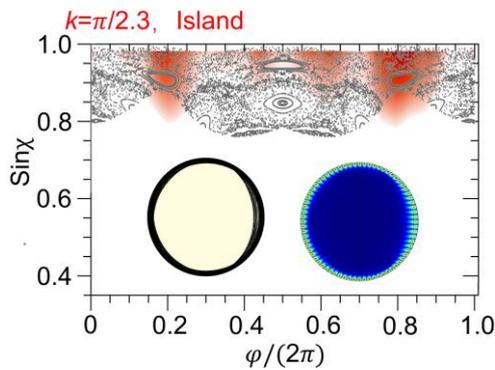

**Figure 4** Husimi projection of the 4-island mode under a very large space curvature of $k=\pi/2.3$. The ray trajectory and spatial field distribution of the mode are included as insets.

Finally, it should be noted that the islands do not provide invariant stability with the varying space curvature. In fact, when the space curvature increases to very large values, the squeezed islands finally break into part of the chaotic sea. This is demonstrated by increasing *k* to $\pi/2.3$, with the corresponding PSOS, Husimi projection, ray trajectory and field distribution of the 4-period island mode [deep red squares in Fig. 1 (d)] shown in Figure 4. The Husimi wavepacket of this island mode exhibits very interesting features. With the vanishing of the 4-period islands, the Husimi wavepacket has to adapt its distribution to the spatial positions of the 3-period islands which are still existing in the PSOS, while nevertheless keeping a four-lobe feature with four intensity maxima. Such a special hybridization of the wave-like behavior is again a unique effect of the curved space.

## 5. Conclusions

In conclusion, we show explicitly in an asymmetric microcavity system that the regular and chaotic motions react to space curvature in very different ways, with both classical and quantum analogies. While the regular modes follow the rise of the phase diagram, the chaotic modes display a special interaction with the rising phase diagram, featured by the effects of localization and hybridization, which uniquely result from the interaction between the wave nature of light and the space curvature. These findings would bring broad interest to mathematics and physics. Apparently, the underlying mechanism for the chaotic behavior vs. space curvature constitutes valuable problems for the mathematicians working in classical and quantum chaos. On the other hand, the study provides an efficient way to tailor optical field properties taking the advantage of the extra degree of freedom associated with the space curvature, utilizing the different behaviors of regular and chaotic motions. To carry on experimental studies in the future, the curved surface can be conveniently fabricated for microwave dynamics, and can be equivalently made as refractive-index-modulated flat structure via transformation optics, functioning at optical wavelengths as photonic chips [35,40]. In return, such photonic chips can be applied in the scenarios of optical communication and computing in which highly flexible tunings of mode frequencies, lifetimes and field distributions are simultaneously required, and can serve as versatile optical simulators for quantum chaotic dynamics and cosmology [41-43].

*This research was supported by National Key R&D Program of China (Grant No. 2023YFA1407100), the National Natural Science Foundation of China (NNSFC) (Grant Nos. 12074303 and 11804267), and Shaanxi Key Science and Technology Innovation Team Project (Grant No. 2021TD-56). F.L. acknowledges the Xiaomi Young Scholar Program.*

**Conflict of Interest**  The authors declare that they have no conflict of interest.